\documentstyle[psfig,epsf]{l-aa}
\input psfig.sty
\def\lsim{\lower.5ex\hbox{$\; \buildrel < \over \sim \;$}}
\def\gsim{\lower.5ex\hbox{$\; \buildrel > \over \sim \;$}} 
\newcommand{\eqb}{\begin{eqnarray}}
\newcommand{\eqe}{\end{eqnarray}}

\begin{document}
\thesaurus{...}
\title{Observational Signature of the `Boundary Layer' of Galactic and
Extragalactic Black Holes} 

\author{S. K. Chakrabarti\inst{1}, L. Titarchuk\inst{2} 
D. Kazanas\inst{3}, K. Ebisawa\inst{3}}
\institute{ Code 665, Goddard Space Flight Center and TIFR, Homi Bhabha Road,
Bombay, 400005
\and
Code 668, NASA/GSFC and George Mason University/CSI
\and
Code 665, Goddard Space Flight Center}

\offprints{S. K. Chakrabarti at TIFR address}
\date{Received \dots ; accepted \dots}
\maketitle
\markboth{Boundary Layers of Black Holes}{}
\begin{abstract}

We present spectral properties of an accretion disk model, in which a
Keplerian accretion disk is flanked by a sub-Keplerian halo component 
terminating at a standing shock. The post-shock region (which may be 
considered to be the boundary layer of a black hole) reprocesses the soft
photons emitted from the Keplerian accretion disk.
We show that switching of states (from hard to soft and  vice
versa) could be accomplished by a change in the accretion
rate of the Keplerian disk component. Our
consideration, for the first time, resolves a long-standing problem of identifying the
illusive `Compton cloud' responsible for the switching of states.

\end{abstract}

Typical continuum spectra of accreting 
galactic and extragalactic black hole candidates exhibit both
`soft' and `hard' components. Occasionally, the same object shows
variabilities in these components such as hardening of the hard component
as the soft luminosity diminishes and vice versa. 
In the former case, the energy spectral index is typically
$0.5 - 0.7$ and the object is considered to be in the 
hard state, whereas in the latter case, the index is typically
$1.2-1.5$ and the object is considered to be in the soft state.
Whereas the hard component is long understood to be the 
characteristic of Comptonization of disk soft photons by an external 
`hot coronae' (as the deficit of soft photons reduces the efficiency 
of cooling of the hot region: Sunyaev \& Titarchuk, 1980, 1985; hereafter
referred to as ST80, ST85 respectively; Titarchuk, 1994; hereafter T94), 
the origin and placement of this hot component have eluded 
satisfactory explanation. We propose that these varied behaviors
could be explained by assuming the supplied matter to be not
strictly Keplerian, as is usually assumed, but a mixture of the
Keplerian matter with a sub-Keplerian halo. The sub-Keplerian matter
could form out of Keplerian disk itself and also be supplemented 
by winds from the companion. Since angular momentum of matter
which could be accreted `with ease' is very small, low viscosity 
flow takes a longer distance to accomplish small angular momentum
at the inner edge, if they start from a Keplerian disk far away.
Higher viscosity flow only passes through inner sonic point, lower viscosity
flow can pass through both.
(Chakrabarti 1990, hereafter C90; Chakrabarti \& Molteni
1995, hereafter CM95; Chakrabarti \& Titarchuk 1995, hereafter CT95;
Chakrabarti 1996, hereafter C96). 
In AGNs, entire matter supply could be sub-Keplerian and a part of 
this can become Keplerian if the viscosity is high enough. 
A generic accretion disk close to a black hole 
would thus have a quasi-Keplerian optically thick disk
in the equatorial plane which is flanked by a sub-Keplerian
halo terminating in an angular momentum supported,
stable, enhanced density region (shock) close to the black hole
horizon. Here the halo `feels' the centrifugal 
barrier and matter piles up behind it (typically, at $r=8-10 r_g$,
$r_g$ is the Schwarzschild radius) for marginally 
bound angular momentum of a Schwarzschild black hole and at a distance
roughly half as much for a rapidly rotating Kerr black hole.
Distance could be much higher if the angular momentum of the
halo component is high (C90). Soft photons radiated by the pre-shock, 
optically thick, geometrically thin, Keplerian flow are intercepted
by the geometrically thick, optically slim ($\tau \sim 1$), 
almost freely falling, hotter post-shock flow and are re-emitted 
as the hard X-ray component after inverse Comptonization. 
In this self-consistent scenario, outflow generated by the
evaporated disk matter and from the companion winds
may become responsible for producing the so-called `reflected' 
component and the observed iron lines which clearly show 
a combination of P-Cygni type and down-scattered type line profile (CT95). 
As discussed in great detail in C90, CT95, CM95, 
C96), the location where the flow deviates from the Keplerian disk
($r_{Kep}$) depends on viscosity parameter and Mach number of the flow, 
the latter being a function of the cooling processes. If the 
cooling efficiency is everywhere negligible (which is valid for
very low accretion rate), then it may be 
impossible to form a Keplerian disk in the first place, in which case,
the Keplerian component on the equatorial plane will be present only very far
away from the black hole. This will then be the quiescence
state of a novae before the outburst. The general picture of how the matter
accretes on a black hole is schematically shown in Fig. 1.

\vspace{2in}

{\small \noindent Fig. 1: Schematic diagram of the most general solution
of the accretion process around a black hole. Keplerian disk which 
produces the soft component is flanked by the sub-Keplerian halo 
which terminates in a standing shock.
Post-shock flow heats up soft photons from the accretion disk
through Comptonization and radiates them as the hard component. Outflows 
from the disk can be responsible for the so-called reflected component
and the Fe emission lines at various stages of ionization.}

Let $L_{I}=f_{ds} L_{SS}$ denote the fraction of the Keplerian 
(i.e., Shakura-Sunyaev [1973] type) disk luminosity $L_{SS}$ of the 
soft radiation intercepted by the bulge of the shock,
$F_e=L_{H}/L_I$ denote the enhancement factor of this flux due to cooling
of the electrons through inverse Comptonization (ST80, ST85, T94)
($L_{H}$ is the hard component luminosity),
and $f_{sd}$ ($\sim 0.25$ for a spherical bulge) denote the
fraction of $L_{H}$ intercepted back by the disk.
The soft component observed from a disk around the black hole candidate is 
therefore contributed by the original disk radiation plus the absorbed
intercepted radiation: $L_{S}=L_{SS} + L_{SS} f_{ds} F_e f_{sd} {\cal B}$,
and $L_{H} \sim L_{SS} f_{ds} F_e$, where ${\cal B}=1-{\cal A}$, 
${\cal A}$ being the albedo of the Keplerian disk.
The enhancement factor $F_e \sim 3(T_{e}/3T_{d})^{1-\alpha} \sim 10-30$
because, typically,
the electron temperature $T_{e} \sim 50$ keV and the disk temperature
$T_d \sim 5$eV for parameters of active galaxies and $T_{e} \sim 150$keV
and $T_d \sim 100$eV for stellar black hole candidates. Hence, we easily 
achieve a convergence, $f_{ds} F_e f_{sd} {\cal B} < 1$ of our algorithm 
since $f_{sd}\sim 0.25$, $f_{ds} \sim 0.05$, and ${\cal B} \sim 0.5$.
Here, $\alpha$ is the energy index ($F_{\nu} \sim \nu^{-\alpha}$)
in the the up-scattering dominated region of the hard spectra.

We now present the detailed continuum spectra resulting 
from the  above considerations (Fig. 1). We included
hydrodynamics of standing shock waves, most accurate
prescription of Comptonization (T94),
the Coulomb exchange of energy between the protons and electrons,
the bremsstrahlung energy loss of the electrons,
and an accurate prescription for the disk albedo.
Two temperature hydrodynamic equations are exactly solved by fourth
order Runge-Kutta method to derive the temperature distributions from 
which the spectral index is computed (see CT95 for detail. N.B.:
the accretion rate sequence in the caption of Fig. 2 of CT95 is 
inadvertantly reversed.).

% GNUPLOT: LaTeX picture with Postscript
\setlength{\unitlength}{0.1bp}
\begin{picture}(2880,1728)(150,0)
\put(1648,1677){\makebox(0,0){ }}
\put(1400,51){\makebox(0,0){$log(\nu)$ }}
\put(100,914){%
%
\makebox(170,920)[b]{\shortstack{$log[\nu F(\nu)]$ }}%
%
}
\put(2447,151){\makebox(0,0){21}}
\put(2147,151){\makebox(0,0){20}}
\put(1848,151){\makebox(0,0){19}}
\put(1548,151){\makebox(0,0){18}}
\put(1249,151){\makebox(0,0){17}}
\put(949,151){\makebox(0,0){16}}
\put(650,151){\makebox(0,0){15}}
\put(350,151){\makebox(0,0){14}}
\put(290,1577){\makebox(0,0)[r]{40}}
\put(290,1312){\makebox(0,0)[r]{38}}
\put(290,1047){\makebox(0,0)[r]{36}}
\put(290,781){\makebox(0,0)[r]{34}}
\put(290,516){\makebox(0,0)[r]{32}}
\put(290,251){\makebox(0,0)[r]{30}}
\end{picture}

{\small \noindent Fig. 2: Variation of spectrum with accretion rate:
${\dot m}_d =1.0$ (dotted), ${\dot m}_d=0.1$ (short
dashed), ${\dot m}_d=0.01$, (long dashed) and ${\dot m}_d=0.001$ (solid). The
accretion rate in sub-Keplerian  halo component is kept fixed at 
${\dot m}_h =1.0$. Hard component softens and soft component 
brightens with ${\dot m}_d$. $M = 5 M_\odot$ is used. Dot-dashed curve 
is drawn when the effects of bulk-motion Comptonization is also added.}

Figure 2 shows a comparison of four runs for the spectra around
a black hole of mass $5 M_\odot$ (uncorrected for the spectral 
gardening factor) and the halo
accretion rate ${\dot m}_h=\frac{{\dot M}_h}{{\dot M}_{Edd}} =1.0$. 
The disk accretion rates ${\dot m}_d=\frac{{\dot M}_d}{{\dot M}_{Edd}}$
are $1.0$ (dotted), $0.1$ (short dashed), $0.01$ (long-dashed) and
$0.001$ (solid). With the increase of the disk accretion rate, soft photons
intercepted by the post-shock bulge is increased, cooling this region
efficiently. Thus the temperature $T_{e}$ of the electrons 
is reduced and the energy index $\alpha$ 
is increased. The luminosity and the peak frequency of the 
soft component go up monotonically with ${\dot m}_d$. The dash-dotted
curve shows the appearance of the weak hard tail due to
bulk-motion Comptonization (Blandford \& Payne, 1981; Titarchuk,
Mastichiadis \& Kylafis, 1996; hereafter TMK96) in the soft state.
When the accretion rate of the disk is very high, the abundant
soft photons cool the post-shock region completely, and the optically
thick, quasi-radial flow (made up of the disk 
and the halo) drags photons along with it, while Comptonizing
them due to the rapid bulk motion prevailing close to the black hole.
Decent fits of soft states of LMC X-3 (Ebisawa, Titarchuk \& Chakrabarti, 
1996; hereafter ETC96) and GS2000-25  and Novae Muscae (Chakrabarti, 1997)
are obtained already. These show that a significant
amount of accretion takes place in the form of sub-Keplerian flow.
In contrast, radial velocity in the
neutron star boundary layer is very small, and the bulk motion
Comptonization  is negligible.

In Fig. 3 we present the energy spectral index $\alpha$ 
(observed slope in the $2-50$keV region) as a function of
the mass accretion rate of the Keplerian disk 
component (${\dot m_{d}}$). Different curves are parameterized (marked) by
the mass accretion rate of the halo component (${\dot m_{h}}$).
The lower left corner represents the so-called `hard-state'  (HS)
with $\alpha \lsim 1$, while middle region represents the so-called 
`soft-state' (SS), $\alpha \gsim 1$. Both are due to reprocessing in the
post-shock flows (PSF). For even high accretion rate on the right, we show 
the spectra from the consideration of (Newtonian) convergent inflow in 
spherical geometry (TMK96) assuming that half the photons are lost at the inner
edge $r=1$ of the flow. For a given halo rate, we note that a transition 
between states can be achieved by a change in ${{\dot m}_d}$.
The important point to note is that both in the extreme hard state
(thermal Comptonization) and the extreme soft state (bulk motion
Comptonization), the spectral index remains almost constant even 
when the accretion rate of the disk changes by orders of magnitude.
This is completely consistent with the observations of the black hole
candidates. In the intermediate accretion rate ($m_d\sim 0.1-1.0$), 
both types of Comptonization may cause double breaks in the hard component.

% GNUPLOT: LaTeX picture with Postscript
\setlength{\unitlength}{0.1bp}
\begin{picture}(2880,1728)(150,0)
\put(1641,700){\makebox(0,0){2.0}}
\put(1585,1030){\makebox(0,0){1.0}}
\put(1538,1480){\makebox(0,0){0.5}}
\put(2300,936){\makebox(0,0){(CF)}}
\put(833,1200){\makebox(0,0)[l]{SS (PSF)}}
\put(2300,1052){\makebox(0,0){SS}}
\put(833,384){\makebox(0,0)[l]{HS (PSF)}}
\put(1648,1677){\makebox(0,0){ }}
\put(1400,51){\makebox(0,0){log(${\dot m}_d$) }}
\put(100,914){%
%
\makebox(170,914)[b]{\shortstack{$\alpha$ }}%
%
}
\put(2447,151){\makebox(0,0){1.5}}
\put(2214,151){\makebox(0,0){1}}
\put(1981,151){\makebox(0,0){0.5}}
\put(1748,151){\makebox(0,0){0}}
\put(1515,151){\makebox(0,0){-0.5}}
\put(1282,151){\makebox(0,0){-1}}
\put(1049,151){\makebox(0,0){-1.5}}
\put(816,151){\makebox(0,0){-2}}
\put(583,151){\makebox(0,0){-2.5}}
\put(350,151){\makebox(0,0){-3}}
\put(290,1577){\makebox(0,0)[r]{3}}
\put(290,1356){\makebox(0,0)[r]{2.5}}
\put(290,1135){\makebox(0,0)[r]{2}}
\put(290,914){\makebox(0,0)[r]{1.5}}
\put(290,693){\makebox(0,0)[r]{1}}
\put(290,472){\makebox(0,0)[r]{0.5}}
\put(290,251){\makebox(0,0)[r]{0}}
\end{picture}

{\small \noindent Fig. 3: Variation of energy spectral index $\alpha$ 
as accretion rate of the disk ${\dot m}_d$ (X-axis) and the halo
${\dot m}_h$ (as marked) are changed. In the left part a hard 
component is produced purely due to the thermal Comptonization, 
whereas in the right part bulk motion Comptonization becomes important. 
In the intermediate region both the effects could be seen (dashed and solid 
curves).}
 
We summarize these properties of our solution
in Table 1 where the correlation (arrow-up) or anti-correlation (arrow-down)
of the observable quantities (luminosities $L_{X,\gamma}$
in X-ray and $\gamma$-ray regions) with the input accretion rates
are shown. Smaller arrow represents a weaker correlation. For a massive 
black hole, results remain very similar as the 
electron temperature is found to be very weak function of the central mass
($T_e \propto M^{0.04-0.1}$).
Shock waves other than those discussed here (e.g., those produced by pre-heating
and magnetic braking), if present, should produce exactly the same type of 
spectra. If the angular momentum of the sub-Keplerian halo is so small that 
shocks do not form at all, or the viscosity and the accretion 
rate (cooling) so small that the entire flow is sub-Keplerian near the
black hole (C90, C96) then the sub-Keplerian component
could intercept the soft photons from the Keplerian disk in the same
way, and switching of states is possible by varying $r_{Kep}$ through viscosity.
However, in this single component scenario there should
always be anti-correlation between the hard and soft states which is 
not observed. We know of no solution other than ours which is based on 
{\it actual} mathematical properties of transonic accretion 
flows on a black hole. After our presentation of this work at the
symposium, some parametric model (with $r_{Kep}$ as an 
{\it adhoc} parameter) has been proposed (Lasota et al, preprint). 
Such one component model should generally produce soft states
(Chakrabarti, 1997).

\medskip

\centerline { TABLE 1}

\medskip

\centerline{ Trends of Luminosity and Spectral Index } 

\begin{center}
\begin{tabular}{|l|l|l|l|}
\hline
    & & & \\
\sl Input & \sl $L_s$ & \sl $L_x,\ \alpha$ & \sl $L_\gamma, \ \alpha$ \\
 \&  & &  & \\
\sl Output & & & \\
\hline
    & & & \\
\sl ${\dot m}_d$ & {\Large $\uparrow$} & {\Large $\uparrow$}
{\Large $\uparrow$}$^a$   & {\Large $\downarrow$}{\Large $\uparrow$}$^a$ \\
    & & & \\
\hline
    & & & \\
\sl ${\dot m}_h$ &  {\small $\uparrow$}
& {\Large $\uparrow$}{\Large $\downarrow$}$^{b,c}$ & 
{\Large $\uparrow$}{\Large $\downarrow$}$^b$ \\
    & & & \\
\hline 
\end{tabular}
\end{center}

\noindent  $^a$ {\small dependence is weaker for ${\dot m}_d \lsim 0.1$ }\\
\noindent  $^b$ {\small dependence is weaker for ${\dot m}_h \gsim 1$ }\\
\noindent  $^c$ {\small $\alpha_X \sim 1 \rightarrow 1.5$, $L_X/L_S \leq 
10^{-3}$, converging flow with very high accretion rate.}

The contribution of the hard component reflected from the accretion disk 
is found at the best to be only ten percent (CT95). The resulting
equivalent width is only around several tens of electron volts.
Since the post-shock region behaves like thick accretion disks 
(Paczy\'nski \& Wiita, 1980; Rees et al., 1982; C96), the rotating outflow 
(Chakrabarti, 1986) can be produced from the funnel of the post-shock region 
which is cooler (because of rotation) and at the same time, has
a large covering factor. Hard photons passing through this
wind will be down scattered to produce the typical iron line emission
spectra (elongated red wing with some or negligible blue absorption)
that is observed. The down scattered X-ray continuum passing 
through this wind will also produce a `transmission bump' commonly 
interpreted as the reflection bump (CT95). However one requires `blobby'
jets to produce the observed equivalent width. Such blobbiness
is indeed observed in disk-jet simulations (Ryu, Molteni, Chakrabarti,
in preparation).

Our present solution of the problem explains shapes of the hard states
from all the black hole candidates, such as {\it GX339-4}, {\it GS2023+338} and
{\it GS1124-68} (Tanaka, Y., 1989; Ueda et al. 1994; Ebisawa et al. 1994).
The black hole candidate novae, {\it GS2000+25, GS2023+338}, and 
{\it GS1124-68} show very similar decay of luminosity
in the post-outburst phase, but the spectral
evolutions are very different. These differences could 
be accounted for by variations of the properties of the halo. For instance,
{\it GS2023+338} is always in hard state throughout the outburst, suggesting a 
high ${\dot m}_h $ but lower ${\dot m}_d $ (because the
black hole mass itself could be higher). The suppression of the soft component
could also be due to a shock distant from the black hole (very high
angular momentum inflow) or very weakly viscous, low accretion
rate, sub-Keplerian flow which intercepts
soft photons from outer Keplerian disk. {\it GS2000+25}, on the other hand,
remained in the soft state during the outburst, suggesting 
a high ${\dot m}_d$, but a low ${\dot m}_h$. In the rising phase of 
{\it GS1124-68} the increase of $L_{H}$ below $\sim 10$ keV is accompanied 
by its decrease above $\sim 10$ keV. This object (along with other black hole
candidates, such as A0620-00) is also
observed in two distinctly different states, as are other black hole
candidates, such as {\it Cyg X-1} and {\it GX339-4}. These are signatures
of increasing accretion rates. A possible scenario of novae outburst
is presented in ETC96.

From Fig. 3 we note that while increase of the spectral slope
requires increasing the disk accretion rate, the same is 
accomplished by {\it decreasing} the halo accretion rate. This
opposite behavior explains occasional scatter in spectral
slope even when the soft luminosity may be increasing (Yaqoob, 1992).
Another important point to note that our results have been plotted
using the dimensionless accretion rates. Thus even if two black hole
candidates show same luminosity, one may be in hard  state while the other
in the soft state if their masses are different. A third point to note
is that, since the location $r_{Kep}$
depends on the accretion rate and viscosity, quiescence states of
black hole candidate novae could be readily explained by a high $r_{Kep}$.
This inner edge comes closer to the black hole
when the viscosity and the Keplerian disk accretion rate becomes higher (C90).
This will produce the novae outbursts (ETC96). In order that this mechanism
be successful, the limit cycle behavior at the Keplerian component
must regulate the disk accretion rate as in the conventional
models (Meyer \& Meyer-Hofmeister, 1989; Cannizzo, 1993).
What we have shown is that hard state or soft state formation
depends on the relative magnitude of the two accretion rates. Thus,
for a given total rate at the outer Keplerian disk, the rates can
redistribute as viscosity changes, and induce corresponding state change.

Although the spectral index of the convergent inflow plotted in Fig. 3
is for spherical flow with zero angular momentum, it is easy to generalize the
result for finite angular momentum (Titarchuk \& Chakrabarti, in preparation).
Fig. 4 shows the variation of spectral index with accretion rate
and angular momentum of the flow in the post-shock region. The solid, dotted,
dashed, long dashed and dot-dashed curves are for $l=0,\ 0.5,\ 1.0,\ 1.5, 2.0$
respectively (in this unit, marginally stable and marginally bound angular
momenta in Schwarzschild geometry are $1.83$ and $2.0$ respectively). 
Observation of $\alpha \sim 1.0-1.5$ along with optical depths of 
the so-called Compton cloud (the post-shock region) $\tau\sim 1-2$ are 
consistent.

\psfig{figure=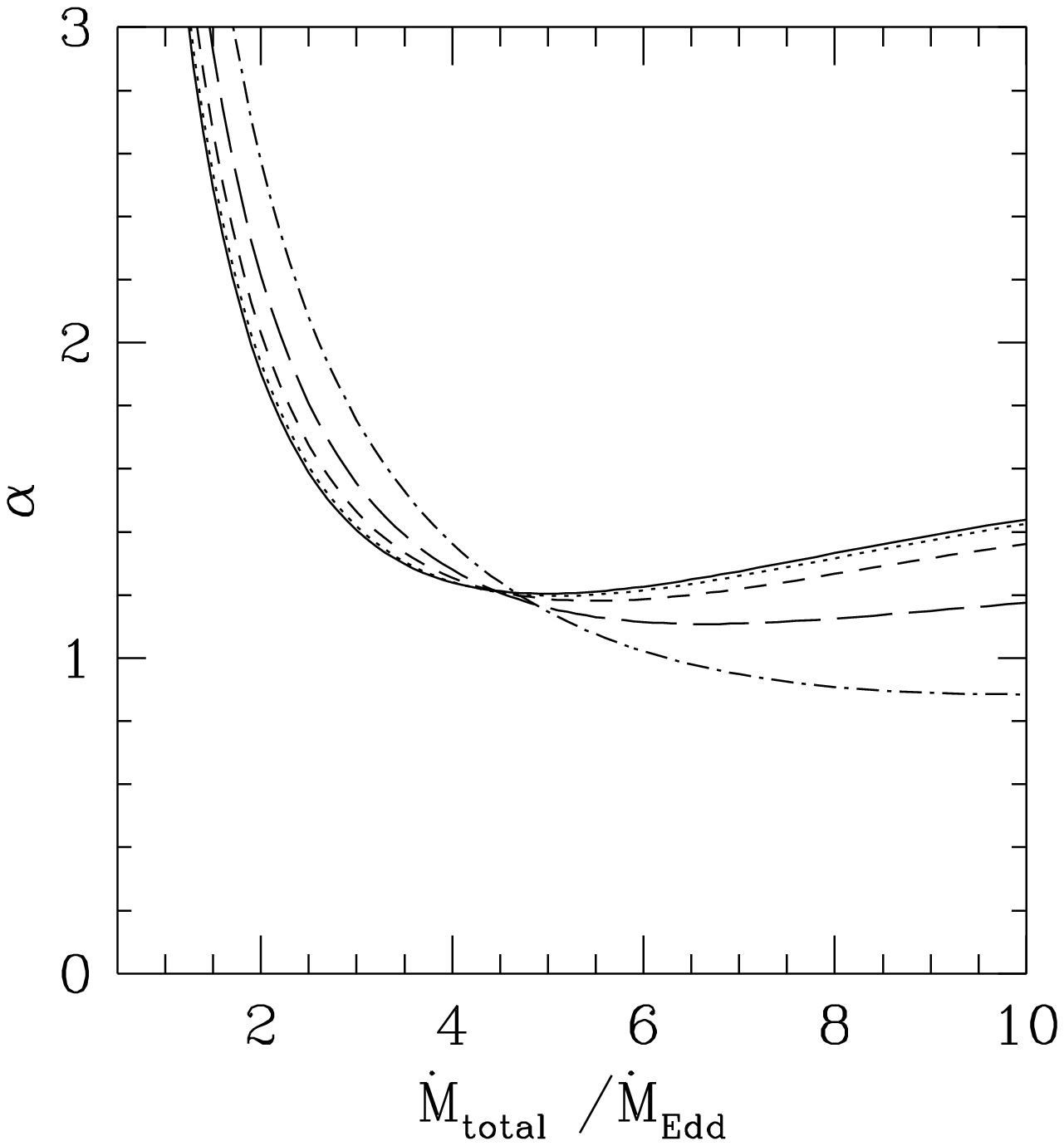,height=2.5in,width=4.0in,rheight=2.0in}

{\small \noindent Fig. 4:  Variation of spectral index in the soft state as 
a function of the net accretion rate in the {\it rotating converging inflow}. 
See text for details.}

{\sl Acknowledgments:} 
Works of SKC and LGT are partially supported by National Research Council
and NASA. The authors thank Tim Kallmann and Tahir Yaqoob for discussions.

{}
\end{document}